\documentclass{elsart}
\newcommand{\nd}{\noindent}
\newcommand{\be}{\begin{equation}}
\newcommand{\ee}{\end{equation}}
\newcommand{\ben}{\begin{eqnarray}}
\newcommand{\een}{\end{eqnarray}}

\usepackage{amssymb}
\usepackage[pctex32]{graphics}
\begin{document}
\begin{frontmatter}


\title{Generating statistical distributions without maximizing the entropy}
\author[plata]{A. Plastino\corauthref{cor}}\corauth[cor]{Corresponding author.}
\ead{plastino@fisica.unlp.edu.ar}  $\,\,$   and
\author[cbpf]{E. M. F. Curado}
\ead{evaldo@cbpf.br}

\address[plata]{Facultad de Ciencias
Exactas, Universidad Nacional de La Plata\\ IFLP-CONICET,
Argentine National
                Research Council\\
                 C.C. 727, 1900 La Plata, Argentina}

\address[cbpf]{Centro Brasileiro de Pesquisas Fisicas (CBPF)\\ Rua Xavier Sigaud 150
- Urca - Rio de Janeiro - Brasil}






\begin{abstract}
 \nd
We show here how to use pieces of thermodynamics' first law  to
generate probability distributions for generalized ensembles when
only  level-population changes are involved. Such microstate
occupation  modifications, if properly constrained via first law
ingredients, can be associated not exclusively  to heat and
acquire a more general meaning.

PACS: 2.50.-r, 3.67.-a, 05.30.-d

 \end{abstract}
\end{frontmatter}
\newpage

\section{Introduction}

 \nd The first  law of thermodynamics is one of physics' most important statements.
Together with the second law, the two constitute strong pillars of
our understanding of Nature. In statistical mechanics an
underlying microscopic substratum is added that is able to explain
not only these laws but the whole of thermodynamics itself
\cite{patria,reif,sakurai,katz}, one of whose
 basic ingredients is a microscopic probability distribution (PD) that
controls the population of microstates of the system under
consideration \cite{patria}. We will be  concerned here {\sf only}
with changes that affect exclusively microstate-population.
 \nd The way  these changes are related to
 changes in a system's extensive quantities provides the  essential
content  of the first law \cite{reif}.
 \nd In this effort we show that the above mentioned  PD
establishes a link between this aspect of the first law, on the
one hand, and the Maximum Entropy principle (MaxEnt), on the other
one,
 according to the
 scheme given below. \begin{itemize} \item \textsc{Hypothesis}: for  \item
a given a concave entropic form (or information measure)~$S$
together with \item 1)  a mean internal energy $U$, 2) mean values
$A_\nu \equiv \langle \mathcal{A}_\nu
\rangle;\,\,(\nu=1,\ldots,M)$ of $M$ additional extensive
quantities $\mathcal{A}_\nu$,  and 3) a temperature $T$, \item
\textsc{Thesis}: then,  for any system described~by
\begin{enumerate} \item
      a microscopic probability
distribution (PD) $\{p_i\}$, and \item
    assuming a {\sf reversible process via} $p_i \rightarrow p_i+dp_i$,
    \item one can verify that:   \newline
 ({\bf 1}) if the  PD $\{p_i\}$ maximizes $S$ this entails
$dU=TdS-\sum_{\nu=1}^M\,\gamma_\nu\,dA_\nu$, {\it or,
alternatively, \newline
 ({\bf 2}) if $dU=TdS-\sum_{\nu=1}^M\,\gamma_\nu\,dA_\nu$,
   this predetermines a unique PD that maximizes}~$S$.
   \end{enumerate} \end{itemize} \nd It should be remarked that, curiously enough,
  this uniqueness of the PD
 does not demand (at this stage) concavity (or convexity) of the entropy
with regard to the distribution of probabilities, a requirement
that arises a posteriori, in further developing the theory of
statistical mechanics \cite{patria}. The transit from ({\bf 1}) to
({\bf 2}) has been studied, for instance,
 in \cite{m1,e1} (by no
means an exhaustive list!).
\nd Succinctly, given a specific  $S-$form, \newline  \newline
$dU=TdS-\sum_{\nu=1}^M\,\gamma_\nu\,dA_\nu\,\,\Leftrightarrow\,\,{\rm
MaxEnt\,\, prob.\,\, distr.\,\, \{p_i\}}.$ \newline

\section{The proof} \nd
 Consider a rather  {\it general} information measure of the form \be
 S= k\sum_i\,p_i\,f(p_i), \ee where, for simplicity's sake,
Boltzmann's constant is denoted just by $k$. The sum runs over a
set of quantum numbers, collectively denoted by $i$
(characterizing levels of energy $\epsilon_i$), that specify an
appropriate basis in Hilbert's space and $\mathcal{P}=\{p_i \}$ is
an (as yet unknown) un-normalized probability distribution such
that $ \sum_i\,p_i=$ constant, the ``constant" being set
eventually equal to unity
(often it is preferably, for practical purposes, to postpone
normalization until the pertinent computation is finished).

Let  $f$ be an arbitrary smooth function of the $p_i$, in such a
way  $p_i f(p_i)$ is a concave function. Further, consider $M$
quantities  $A_\nu$ that represent  mean values of extensive
physical quantities $\mathcal{A}_\nu$. These take, for the state
$i$, the value
  $a_i^\nu$ with probability $p_i$.
 Also, we suppose that $g$ is
another arbitrary  smooth, monotonic function of the $p_i$ such
that  $g(0) = 0$ and  $g(1) = 1$. We do not need to require the
condition $\sum_i g(p_i) = 1.$
The mean energy $U$ and the $A_\nu$ are given by \ben \label{4} U=
\sum_i\,\epsilon_i\,g(p_i); \,\,\, A_\nu =
\sum_i\,a_i^\nu\,g(p_i). \een \nd Assume now that the
probability-set $\mathcal{P}$ changes in the fashion \be \label{5}
p_i \rightarrow p_i \,+\, dp_i,\,{\rm
with}\,\,\sum_i\,dp_i=0\,\,{\rm (normalization!)},\ee which in
turn generates corresponding changes $dS$, $d A_\nu$, and $dU$ in,
respectively, $S$, the $A_\nu$, and $U$.

 \nd The essential point that we are introducing in this effort is
that we want  {\it to make sure that, in the above described
circumstances, the following condition, related to the first law,
is obeyed} \be \label{6} dU-TdS + \sum_{\nu=1}^M dA_\nu
\lambda_\nu=0, \ee with $T$ the temperature. As a consequence of
(\ref{6}), a little algebra yields, up to first order in the
$dp_i$, the
condition  
\ben \label{777} & \sum_i [C^{(1)}_i + C^{(2)}_i] dp_i \equiv \sum
K_i dp_i = 0; \cr &  C^{(1)}_i= [\sum_{\nu =1}^M
\lambda_\nu\,a_i^\nu +\epsilon_i]\,g'(p_i);  \,\,\,  C^{(2)}_i=
-kT [f(p_i)\,+\,p_i\,f'(p_i)], \een where the primes indicate
derivative with respect to $p_i$. Eq. (\ref{777}) should hopefully
yield one and just one expression for the $p_i$.
\subsection{Equality of the $K_i$ in Eq. (\ref{777})}
\nd We proceed to show now that all the $K_i$ are equal. As the
$dp_i$ are linked by the relation \be \sum_i\,K_i\, dp_i = 0,
\label{firste} \ee we can write, if we are dealing with  $N$
micro-states, \be \label{seconde} dp_N = -\sum^{N-1}_{i=1}\, dp_i.
\ee Substituting (\ref{seconde}) in (\ref{firste}) we obtain \be
\label{thirde}   \sum^{N-1}_{i=1}\,(K_i - K_N)\,dp_i = 0. \ee Now,
since the $N - 1$ ``population-variations" $dp_i$ are independent,
this entails  that each term  in (\ref{thirde})  should vanish by
itself, which implies that \be \label{fourthe} K_i = K_N;\,\,\,
{\rm for \,\,\, all\,\,\,} i = 1,\ldots,N - 1.\ee Thus, $K_i=$
constant  (in the sense of being independent of $i$) $=K$ for all
$i$.

\subsection{The role of $K$}

\nd Interestingly enough, we do not need to give a specific value
to $K$ for our present purposes, although it will become clear
below that it is related to the probabilities-normalization
constant. We only need to ascertain the $K-$role, in the following
sense. We have, on account of (\ref{fourthe}),
 \ben \label{77} & K= C^{(1)}_i + C^{(2)}_i;\,\,\,(for\,\,\,any\,\,\,i), \cr
 & C^{(1)}_i= [\sum_{\nu =1}^M \lambda_\nu\,a_i^\nu +\epsilon_i]\,g'(p_i),\cr
 & C^{(2)}_i= -kT [f(p_i)\,+\,p_i\,f'(p_i) ], \een
  so that, if  we  redefine things  in the fashion
 \ben \label{7a}
 & T^{(1)}_i = f(p_i)\,+\,p_i\,f'(p_i) \cr
 & T^{(2)}_i= - \beta [(\sum_{\nu =1}^M \lambda_\nu\,a_i^\nu +\epsilon_i)\,g'(p_i) - K],
 \,\,\, (\beta \equiv 1/kT), \een
  we can recast (\ref{77}) as \be \label{7}
 T^{(1)}_i+T^{(2)}_i=0;\,\,\,(for\,\,\,any\,\,\,i),   \ee
a relation whose importance will be presently become manifest.
\subsection{The MaxEnt route revisited}
\nd  Assume now that you wish to extremize $S$ subject to the
constraints of fixed valued for i) $U$ and ii) the $M$ values
$A_\nu$. This is achieved via Lagrange multipliers (1) $\beta$ and
(2) $M$ $\gamma_\nu$. We need also a normalization Lagrange
multiplier~$\xi$. \be  \label{mop} \delta_{\{\ p_i\}}
\left[S-\beta U -\sum_{\nu =1}^M\, \gamma_\nu A_\nu   - \xi
\sum_i\,p_i\right] =0,   \ee leading to, with $\gamma_\nu = \beta
\lambda_\nu$, to \be \label{mep1}
  0= \delta_{p_m} \sum_i \left( p_i f(p_i)
   -  [\sum_i \beta g(p_i)
  (\sum_{\nu =1}^M \lambda_\nu\,a_i^\nu + \epsilon_i) +\xi p_i]\right),   \ee
 so that \ben
  & 0= f(p_i) + p_i f'(p_i) -[ \beta g'(p_i)
 (\sum_{\nu =1}^M \lambda_\nu\,a_i^\nu +\epsilon_i) +\xi ]\Rightarrow
 \,\,{\rm if}\,\,\,\xi\equiv\beta K, \cr
  & 0= f(p_i) + p_i f'(p_i) -  \beta [g'(p_i)
  (\sum_{\nu =1}^M \lambda_\nu a_i^\nu +\epsilon_i) +K] \Rightarrow
  \cr & 0= T^{(1)}_i+T^{(2)}_i. \label{mep} \een
\nd Clearly, (\ref{7}) and the last equality of (\ref{mep}) are
{\sf one and the same equation}! The equivalence stated in the
Abstract is thus proven.

\section{Discussion}

\nd We have here endeavored to show that appropriate manipulation
of some ingredients of the first law of thermodynamics can be used
 to generate the equilibrium   microscopic probability distribution (PD)
that
 describes a system within the framework of a generalized ensemble
\cite{patria}, and that such an approach is an alternative to the
MaxEnt-one .  \nd We were exclusively concerned  with changes that
affect exclusively microstate-population and, more specifically,
with
 the way  these modifications are related to
 internal variation of the system's extensive quantities
 \cite{reif}.
\nd We started with
 (1) a  given a concave entropic form (or an information measure (IM)) $S$,
 (2) a mean internal
energy $U$,
and $M$  mean values
$A_\nu \equiv \langle \mathcal{A}_\nu \rangle;\,\,(\nu=1,\ldots,M)$
of $M$ extensive quantities $\mathcal{A}_\nu$,  (3) a temperature $T$,
 and demonstrated that, for any system described by a microscopic probability
distribution (PD) $\{p_i\}$,
   assuming a reversible process via
   $p_i \rightarrow p_i+dp_i$ that is forced to verify the
   relation
 $dU=TdS-\sum_{\nu=1}^M\,\gamma_\nu\,dA_\nu$, we got an equation that yields
 a unique PD that maximizes
  $S$. By way of contrast, MaxEnt starts from $S$ and,
  extremizing it with appropriate constraints, allow one to find the system's PD.  
\nd In other words,
$dU=TdS-\sum_{\nu=1}^M\,\gamma_\nu\,dA_\nu\,\,\Leftrightarrow\,\,{\rm
MaxEnt\,\, prob.\,\, distr.\,\, \{p_i\}}.$ An alternative route to
microscopic PD's, with some first law flavor,  has thus been found
in the present communication .

  \end{document}